\documentclass[twocolumn,prl,amsmath,amssymb]{revtex4}

\def\be{\begin{equation}}
\def\ee{\end{equation}}
\def\tr{\mbox{tr}}
\def\one{\openone}
\def\compl{\mathbb{C}}
\def\bra#1{\langle#1|} \def\ket#1{|#1\rangle}

\def\proj#1{\ket{#1}\!\bra{#1}}

\begin{document}

\title{ Noise robustness of the nonlocality of entangled quantum states }

\author{Mafalda L. Almeida$^{1}$, Stefano Pironio$^{1}$, Jonathan Barrett$^2$, G\'eza T\'oth $^{1,3}$, and Antonio Ac\'\i n$^{1,4}$}

\affiliation{ $^1$ICFO-Institut de Ciencies Fotoniques, E-08860
Castelldefels, Barcelona, Spain\\
$^2$Perimeter Institute for Theoretical Physics, 31 Caroline Street N, Waterloo, Ontario, Canada N2L 2Y5\\
$^3$Research Institute for Solid State Physics and Optics,
%Hungarian Academy of Sciences,
P.O. Box 49, H-1525 Budapest, Hungary\\
$^4$ICREA-Instituci\'o Catalana de Recerca i Estudis Avan\c{c}ats, Lluis Companys 23, 08010 Barcelona, Spain
 }

\date{\today}

%%%%%%%%%%%% Abstract %%%%%%%%%%%%%%%%%%%%%%%%%%%

\begin{abstract}
We study the nonlocal properties of states resulting from the
mixture of an arbitrary entangled state $\rho$ of two
$d$-dimensional systems and completely depolarized noise, with
respective weights $p$ and $1-p$. We first construct a local model
for the case in which $\rho$ is maximally entangled and $p$ at or
below a certain bound. We then extend the model to arbitrary $\rho$.
Our results provide bounds on the resistance to noise of the
nonlocal correlations of entangled states. For projective
measurements, the critical value of the noise parameter $p$ for
which the state becomes local is at least asymptotically $\log(d)$
larger than the critical value for separability.
\end{abstract}


\maketitle

%\section{Introduction}
In 1964, Bell showed that some entangled states are nonlocal, in the
sense that measurements on them yield outcome correlations that
cannot be reproduced by a locally causal model \cite{Bell}. This
nonlocal character of entangled states may be demonstrated through
the violation of Bell inequalities. All pure entangled states
violate such an inequality, hence are nonlocal \cite{Gisin}. For
noisy states, the picture is much subtler. Werner constructed in
1989 a family of bipartite mixed states, which, while being
entangled, return outcome correlations under projective measurements
that can be described by a local model \cite{Werner}. This result
has been extended to general measurements \cite{Barrett} and more
parties \cite{TA}. Thus, while entanglement is necessary for a state
to be nonlocal, in the case of mixed states it is not sufficient.

Beyond these exploratory results, little is known about the relation
between noise, entanglement, and quantum nonlocality. Understanding
this relation, apart from its fundamental interest, is important
from the perspective of Quantum Information Science. In this
context, entanglement is commonly viewed as a useful resource for
various information-processing tasks. Not all entangled states,
however, are useful for every task: for example, quantum computation
with slightly entangled states can be efficiently simulated on a
classical computer \cite{comp}, and bound entangled states are
useless for teleportation \cite{HHH}. For certain tasks, such as
quantum communication complexity problems \cite{commcompl}, or
device-independent quantum key distribution \cite{AGM}, entangled
states are useful only to the extent that they exhibit nonlocal
correlations. Indeed, in these scenarios two (or more) distant
observers, Alice and Bob, directly exploit the correlations
\begin{equation}\label{qcorr}
    P_{MN}(a,b)=\tr(\rho_{AB}\,M_a\otimes N_b)\,,
\end{equation}
obtained by performing measurements $M$ and $N$ on a distributed
entangled state $\rho_{AB}$ (in the above formula, $M_a$ and $N_b$
are the positive operators associated with the measurement
outcomes $a$ and $b$). If the entangled state $\rho_{AB}$ can be
simulated by a local model, these correlations can be written as
\begin{equation}\label{loccorr}
    P_{MN}(a,b)=\int\!\! \mu(d\lambda)\, P_M(a|\lambda)P_N(b|\lambda)\, ,
\end{equation}
where $\lambda$ denotes a shared classical variable distributed
with probability measure $\mu$, and $P_M(a|\lambda)$ and
$P_N(b|\lambda)$ are the local response functions of Alice and
Bob. For all practical purposes then, the entangled state
$\rho_{AB}$ can be replaced by classical correlations, and so does
not provide any improvement over what is achievable using
classical resources \cite{Lluis}.

In this work, we estimate the resistance to noise of the nonlocal
correlations of bipartite entangled states in $\compl^d\otimes
\compl^d$, where $d$ is the local Hilbert space dimension of each
subspace. To do this, we analyze the nonlocal properties of states
resulting from the mixture of an arbitrary state $\rho$ with
completely depolarized noise,
\begin{equation}\label{noisy_states}
    \rho(p)=p\,\rho+(1-p)\frac{\one}{d^2}\, .
\end{equation}
Our goal is to find the minimal amount of noise that destroys the
nonlocal correlations of any state $\rho$, i.e., the maximal value
$p_L$ such that $\rho(p)$ is local for any $\rho$ when $p\leq p_L$.
Clearly, for sufficiently small values of $p\leq p_S$, the state
$\rho(p)$ becomes separable for any $\rho$ \cite{sep, GB}, thus
local. We give here lower bounds on $p_L$ that are more constraining
than the one obtained from the separability condition. If we
restrict Alice and Bob to perform projective measurements only, the
bound that we obtain for the locality limit is asymptotically
$\log(d)$ larger than the separability limit.

A key step in the proof of our results is the construction of a
local model for states of the form (\ref{noisy_states}) when
$\rho=\ket{\phi_d}\bra{\phi_d}$ is maximally entangled, i.e.
$\ket{\phi_d}=1/\sqrt d\sum_{i=1}^d \ket{ii}$. Thus we also provide
a lower bound on $p_L^{\phi}$, defined as the maximal value of $p$
such that
\begin{equation}
p\,\ket{\phi_d}\bra{\phi_d}+(1-p)\frac{\one}{d^2}
\end{equation}
is local. This last result implies in particular the existence of
entangled states whose nonlocal correlations are more robust than
those of maximally entangled ones.

The results presented here concern mostly the simpler but
physically relevant case in which Alice and Bob are restricted to
projective measurements. Extensions to completely general
measurements are discussed at the end of the paper. Our results
also provide bounds for the notion of state steerability
introduced in \cite{WJD}.

As mentioned, we start by analyzing the case in which the state $\rho$ in
(\ref{noisy_states}) is maximally entangled. %, i.e., we consider the
%states
%\begin{equation} \label{isotr state}
%\rho_I=p\, \proj{\phi_d}+(1-p)\frac{\one}{d^2}\, ,
%\end{equation}
%$\rho=p\, \proj{\phi_d}+(1-p)\one/d^2$.
Such states are
called isotropic states and are the unique ones invariant under
$U\otimes U^*$ transformations for all unitary operators $U$ on
$\compl^d$ \cite{HH}. If Alice and Bob each make on these states a
projective measurement, specified by a set of $d$ orthogonal
projectors $Q=\{Q_a\}$ for Alice and $R=\{R_b\}$ for Bob, with
$a,b=1,\ldots,d$, the resulting joint outcome probabilities are
given by
\begin{equation}
\label{quantum prob proj} \frac{p}{d}\tr\left(Q_a^T
R_b\right)+\frac{1-p}{d^2}\, .
\end{equation}
Our first aim is to construct a local model for isotropic states,
that is, to write the quantum probabilities \eqref{quantum prob
proj} in the form (\ref{loccorr}) for some value of the noise
parameter $p$.

Our construction is inspired by the model given in Ref.
\cite{Werner} for Werner states, which are $U\otimes U$ invariant,
and which we adapt to the $U\otimes U^*$ symmetry of isotropic
states. The local classical variables $\lambda$ in our model are
taken to be complex $d$-dimensional vectors which we can thus
formally identify with $d$-dimensional quantum states
$\ket{\lambda}$. The probability measure $\mu$ is
the unique measure invariant under all unitary transformations $U$
on $\compl^d$. In analogy with the quantum formalism, Alice's
response function is defined as
\begin{equation}
\label{aliceresp} P_Q(a|\lambda)=\bra{\lambda}Q_a^T\ket{\lambda}\,
.
\end{equation}
Bob's response function is suggested by the perfect correlations of maximally entangled states and taken to be
\begin{equation}
\label{bobresp} P_R(b|\lambda)=\begin{cases}
1& \text{if } \bra{\lambda}R_b\ket{\lambda}=\max_{i}{\bra{\lambda}R_i\ket{\lambda}}\\
0& \text{otherwise}\,.
\end{cases}
\end{equation}
It satisfies \be\label{Ubob} P_{U^\dagger R
U}(b|\lambda)=P_R(b|U\lambda)\,. \ee To obtain the joint
probabilities predicted by this model, and to compare them with
\eqref{quantum prob proj}, it is necessary to compute the integral
\eqref{loccorr} for our specific choice of measure $\mu$ and
response functions. Following Werner (see \cite{Werner} for
details), one can show that the $U$-invariance of $\mu$, the form
\eqref{aliceresp} of Alice's response function, and the relation
\eqref{Ubob} satisfied by Bob's response function, imply
\begin{equation}\label{integral LHV}
\int\!\!\mu(d\lambda)\,P_Q(a|\lambda)P_R(b|\lambda)=\tr\left(Q_a^T\hat
B(b,R)\right)\,,
\end{equation}
where $\hat B(b,R)$ is a positive operator depending on Bob's
response function. One can further show, exploiting the fact that
the relation \eqref{integral LHV} holds for all one-dimensional
projectors $Q_a$ \cite{Werner}, that $\hat B(b,R)=(p^\phi/d)\,
R_b+(1-p^\phi)/d^2\, \one$, for some $p^\phi\in \mathbb{R}$, and
thus that
\begin{equation}\label{integral LHV 2}
\int\!\!\mu(d\lambda)P_Q(a|\lambda)P_R(b|\lambda)\!=\!\frac{p^\phi}{d}\tr\left(Q_a^T
R_b \right)+\frac{1-p^\phi}{d^2}\,.
\end{equation}
These correlations are thus already of the prescribed form
(\ref{quantum prob proj}). To determine the value of $p^\phi$ for
which \eqref{integral LHV 2} holds, it is sufficient to compute
the integral \eqref{integral LHV} in the simplest case where
$Q_a^T=R_b$, which gives
\begin{equation}\label{pcr}
p^\phi=\frac{1}{d-1}\left(-1+d^2\int\!\!
\mu(d\lambda)\,\bra{\lambda}R_b
\ket{\lambda}P_R(b|\lambda)\right)\, .
\end{equation}
It now remains to evaluate this integral for the specific choice
(\ref{bobresp}) for $P_R(b|\lambda)$. After patient algebra, one
obtains
\begin{equation}\label{boundmax}
p^\phi=\frac{1}{d-1}\left(-1+\sum_{k=1}^d\frac{1}{k}\right)
\,\xrightarrow[\mathrm{large}\; d]{}\,\frac{\log(d)}{d}\,.
\end{equation}
For $d=2$, $p^\phi=1/2$ is equal to the critical value for
two-dimensional Werner states, as expected since Werner and
isotropic states are equivalent up to local unitary
transformations when $d=2$. In the limit of large $d$, $p^\phi$ is
asymptotically $\log(d)$ larger than the critical probability
$p^\phi_S=1/(d+1)$ for the separability of isotropic states
\cite{HH}.

Our next goal is to generalize the local model for isotropic states
to mixed states of the form
\begin{equation} \label{pure+noise}
\rho=p\,\proj{\psi}+(1-p)\frac{\one}{d^2}\, ,
\end{equation}
where $\ket{\psi}$ is an arbitrary pure state in
$\compl^d\otimes\compl^d$. This automatically also implies a model
for the general states (\ref{noisy_states}), since any mixed state
$\rho$ is a convex combination of pure states. To do this, we
incorporate Nielsen's protocol \cite{nielsen} for the conversion
of bipartite pure states by local operations and classical
communication (LOCC) into our model. Recall that a maximally
entangled state $\ket{\phi_d}$ can be transformed by LOCC in a
deterministic way into an arbitrary state $\ket{\psi}$ by a single
measurement on Alice's particle followed by a unitary operation on
Bob's side, depending on Alice's measurement outcome. Indeed,
consider an arbitrary pure entangled state written in its Schmidt
form $\ket{\psi}=\sum_{j=0}^{d-1}\nu_j\ket{jj}$, and denote by
$D_\nu$ the $d\times d$ diagonal matrix with entries
$(D_\nu)_{jj}=\nu_j$. Taking the $d$ cyclic permutations
$\Pi_i=\sum_{j=0}^{d-1} \ket{j}\bra{j+i \,(\mathrm{mod}\, d)}$,
where $i=0,\ldots,d-1$, it is possible to write
\begin{equation}
\label{nielsenrel} \ket{\psi}=\sqrt {d}(A_i\otimes \Pi_i)
\ket{\phi_d}\quad
\text{for all }
i=0,\ldots,d-1\, ,
\end{equation}
with $A_i=D_\nu\Pi_i$. The operators $W_i=A_i^\dagger A_i$ define a
measurement, since they are positive and sum to the identity,
$\sum_i W_i=\one$. In order to convert $\ket{\phi_d}$ into
$\ket{\psi}$, Alice first carries out this measurement, obtaining
the outcome $i$ with probability $\bra{\phi_d}W_i\ket{\phi_d}=1/d$.
She then communicates her result to Bob who applies the
corresponding unitary operation $\Pi_i$, the resulting normalized
state being $\ket{\psi}$, as implied by (\ref{nielsenrel}).

The quantum-like properties of our local model, i.e., the fact
that the hidden variable $\ket{\lambda}$ can be thought of as a
quantum state and the quantum form of the response function
(\ref{aliceresp}), allow us to adapt Nielsen's construction to it.
The idea is that at the source, before sending the classical
instructions $\ket{\lambda}$ to each party, a measurement defined
by the operators $A^*_i$ is simulated on $\ket{\lambda}$, giving
outcome $i$ with probability $q_i(\lambda)=\bra{\lambda}A_i^T
A_i^*\ket{\lambda}$. The classical description of the normalized
hidden states $\ket{\lambda^A_i}=A^*_i\ket{\lambda}/\sqrt{q_i}$
and $\ket{\lambda^B_i}=\Pi_i\ket{\lambda}$ are then sent,
respectively, to Alice and Bob, who use them in the response
functions \eqref{aliceresp} and \eqref{bobresp} instead of
$\ket{\lambda}$. The joint probabilities $P_{QR}(a,b)$ predicted
by the model for measurements $Q$ and $R$ are thus given by
\begin{multline}\label{nielsenmodel}
 \int\!\! \mu(d\lambda)\, \sum_{i=0}^{d-1} q_i(\lambda) P_Q(a|\lambda^A_i)P_R(b|\lambda^B_i) \\
=\sum_{i=0}^{d-1} \int\!\! \mu(d\lambda)\, \bra{\lambda}A_i^T
Q_a^T A_i^*\ket{\lambda} P_{\Pi^{\dagger}_i R_b \Pi_i}(b|\lambda)
\end{multline}
where we used property \eqref{Ubob}. Replacing the integral in the last expression by the right-hand side of  \eqref{integral LHV 2}, we obtain
\begin{equation}
    \sum_{i=0}^{d-1}\left(\frac{p^\phi}{d}\tr(A_i^T Q_a^TA_i^*\Pi_i^\dagger R_b \Pi_i)+
    \frac{1-p^\phi}{d^2}\tr(A_i^T Q_a^TA_i^*)\right)\, .
\end{equation}
Using Eqs. (\ref{nielsenrel}) and \eqref{quantum prob proj}, and
the fact that $\sum A_iA_i^\dagger=d\sigma$, where
$\sigma=\tr_B\proj{\psi}$, one can check that these probabilities
are equal to the quantum probabilities
$\tr\left(\tilde\rho\,Q_a\otimes R_b \right)$ for the state
\begin{equation}\label{noisy_psi}
    \tilde\rho=p^\phi\proj{\psi}+(1-p^\phi)\sigma\otimes\frac{\one}{d}\, .
\end{equation}
Not surprisingly, the measurement at the source modifies the local noise of Alice, which is no longer completely depolarized, and
introduce some bias depending on $\ket{\psi}$.

This result can already be interpreted as a measure of the
robustness of the nonlocal correlations of an arbitrary entangled
state $\ket{\psi}$. By mixing a state-dependent local noise, with
mixing probability $1-p^\phi$, it is always possible to wash out
the nonlocal correlations of the state $\ket{\psi}$.

In order to extend this result to the case of completely
depolarized noise, one can add some extra local noise to Alice
such that the resulting state has the form \eqref{pure+noise},
with the penalty that $p< p^\phi$. Writing the reduced density
matrix $\sigma$ in its diagonal form $\sigma=\sum_j \mu_j^2
\proj{j}$, and defining $\sigma_k=\sum_j \mu_{j+k
\,(\mathrm{mod}\,d)}^2 \proj{j}$, it is clear that the state
\begin{equation}
\label{mixtonoise}
q\tilde\rho+\frac{1-q}{d-1}\sum_{k=1}^{d-1}\sigma_k\otimes\frac{\one}{d}
\end{equation}
has the form \eqref{pure+noise} for $q(1-p_d)=(1-q)/(d-1)$, in which case the
probability $p$ is given by
\begin{equation}\label{asympt pr}
p^\rho=\frac{p^\phi}{(1-p^\phi)(d-1)+1}\,\xrightarrow[\mathrm{large}\;
d]{}\,\frac{\log(d)}{d^2}\,.
\end{equation}
The state (\ref{mixtonoise}) is clearly local, since it is a
convex combination of local states. We have thus shown that the
noisy states (\ref{noisy_states}) have a local model for
projective measurements whenever $p\leq p^\rho$.  The
probabilities $p^\phi$ and $p^\rho$ represent the main results of
this work and provide lower bounds on $p_L^{\phi}$ and $p_L$.
Several implications of our findings are discussed in what
follows.

First of all, one may ask about the tightness of our bound.
Actually, our model is based on Werner's construction, and this
model is known not to be tight in the case $d=2$ \cite{AGT}. Even
if it is not tight, it would be interesting to understand whether
the model predicts the right asymptotic dependence with the
Hilbert-space dimension $d$. An upper bound on $p_L$ follows from
the results of \cite{ADGL}, where it was shown that a state of the
form $\varrho_2=p \ket{\phi_2}\bra{\phi_2}+(1-p)\one/d^2$, where
$\ket{\phi_2}=1/\sqrt{2}(\ket{00}+\ket{11})$ is a projector onto a
two-qubit maximally entangled state, violates the
Clauser-Horne-Shimony-Holt inequality \cite{CHSH} whenever
$p>p^{\varrho_{2}}$, where
\begin{equation}\label{upprob}
    p^{\varrho_2}=\frac{4(d-1)}{(\sqrt 2-1)d^2+4d-4}\,\xrightarrow[\mathrm{large}\; d]{}\,\frac{4}{(\sqrt 2-1)d}\,,
\end{equation}
which tends to zero when $d\to\infty$. This result together with
our previous model thus imply that $p^\rho\leq p_L\leq
p^{\varrho_2}$.

\begin{table*}[t]
\caption{Asymptotic bounds on the critical noise threshold for separability ($p_S$) and locality ($p_L$) for maximally entangled states ($\ket{\phi_d}$) and arbitrary states ($\rho$). For maximally entangled states, $p_S^{\phi}$ is given in \cite{HH}; the lower bounds for $p_L^\phi$ follow from eqs.~\eqref{boundmax} and \eqref{probpovm}; and the upper bounds from \cite{CGLMP}, where $K$ is Catalan's constant. For arbitrary states, bounds for $p_S$ were derived in  \cite{GB}; the lower bounds for $p_L$ are obtained from those for the maximally entangled states using eq.~\eqref{asympt pr}; the upper-bounds are those of \cite{ADGL}.
}
\begin{ruledtabular}
\begin{tabular}{c|lll}
state & separability & locality (projective meas.) & locality (general meas.)\\
\hline$\ket{\phi_d}$ & $p^\phi_S=\frac{1}{d+1}$ & $\Theta\left(\frac{\log d}{d}\right)\leq p^\phi_L\leq\frac{\pi^2}{16\,K}\simeq 0.67$ & $\Theta\left(\frac{3}{e\,d}\right)\leq  p^\phi_L\ \leq\frac{\pi^2}{16\,K}\simeq 0.67$\\
arbitrary $\rho$ & $\frac{1}{d^2-1}\leq p_S \leq \frac{2}{d^2+2}$ & $\Theta\left(\frac{\log d}{d^2}\right)\leq p_L\leq \Theta\left(\frac{4}{(\sqrt{2}-1)d}\right)$&$\Theta(\frac{3}{e\,d^2})\leq p_L\leq\Theta\left(\frac{4}{(\sqrt{2}-1)d}\right)$
\end{tabular}
\end{ruledtabular}
\end{table*}

Our results, when combined with (\ref{upprob}), also provide a
strict proof of the fact that the nonlocal correlations of
maximally entangled states, under projective measurements, are not
the most robust ones. Indeed, we have a local model for isotropic
states whenever $p\leq p^\phi$, while there exist quantum states
of the form (\ref{noisy_states}) violating a Bell inequality when
$p>p^{\varrho_2}$. For sufficiently large dimension,
$p^{\varrho_2}<p^\phi$, so we have a Bell inequality violation in
a range of $p$ for which we have shown the existence of a local
model for isotropic states.

It is also interesting to compare the bounds derived here for
nonlocality with those known for entanglement. To our knowledge, the
best upper and lower bound on the critical probability $p_S$ such
that the states (\ref{noisy_states}) are guaranteed to be separable were obtained in
Ref. \cite{GB}:
\begin{equation}\label{sep_bounds}
    \frac{1}{d^2-1}\leq p_S\leq \frac{2}{d^2+2}\, .
\end{equation}
Interestingly, the upper bound is obtained, as above, for the case
in which the state $\rho$ in (\ref{noisy_states}) is equal to a
projector onto $\ket{\phi_2}$. Comparing with Eq. (\ref{asympt
pr}), we see that the critical noise probability for nonlocality
under projective measurements is, at least, asymptotically
$\log(d)$ larger than the one for separability, as it is for
isotropic states.

Finally, let us briefly mention how the above results can be
extended to the case of general measurements. The idea is, as
above, to start by constructing a model for isotropic states,
adapting the one for Werner states of Ref. \cite{Barrett}. As
noted in \cite{Barrett}, it is sufficient to simulate measurements
$M$ and $N$ defined by operators $M_a=c_a Q_a$ and $N_b=c_b R_b$
proportional to one-dimensional projectors $Q_a$ and $R_b$ to be
able to simulate any measurement by Alice and Bob. In our
corresponding model, the hidden states are again vectors
$\ket{\lambda}$ in $\compl^{d}$ chosen with the Haar measure
$\mu$. Alice's response function is basically the same as before,
\begin{equation}\label{alicepovm}
    P_M(a|\lambda)=\bra{\lambda}M_a^T\ket{\lambda}\,,
\end{equation}
while Bob's is, taking inspiration from \cite{Barrett}, chosen as
\begin{multline}\label{bobpovm}
  P_N(b|\lambda) = \bra{\lambda}N_b\ket{\lambda}
\Theta\!\left(\bra{\lambda}R_b\ket{\lambda}-\frac{1}{d}\right)\\
   + \frac{c_b}{d}\left[1-\sum_k \bra{\lambda}N_k\ket{\lambda}\Theta\!\left(\bra{\lambda}R_k\ket{\lambda}-
\frac{1}{d}\right)\right]\,,
 \end{multline}
where $\Theta$ is the Heaviside step function. Evaluation of the
integral \eqref{loccorr} with the definitions \eqref{alicepovm}
and \eqref{bobpovm} can be done along the same steps as in
\cite{Barrett} and yields the joint measurement outcome
probabilities for an isotropic state with the critical value
\begin{equation}\label{probpovm}
    \tilde p^\phi=\frac{(3d-1)(d-1)^{d-1}}{(d+1)d^d}\,\xrightarrow[\text{large }d]{}\,\frac{3}{e}\frac{1}{d}\, .
\end{equation}

Since this model has the same quantum-like properties as the one
for projective measurements, cf.~definition~(\ref{alicepovm}), it
can also be extended to arbitrary noisy
states~(\ref{noisy_states}) using Nielsen's protocol. The
corresponding critical probability is given by (\ref{pcr}) with
$p^\phi$ replaced by the above value of $\tilde p^\phi$.

In conclusion, we have obtained bounds on the robustness of the
nonlocal correlations of arbitrary entangled states. Our results
are summarized in Table I.  In the particular but interesting case
where the state is maximally entangled, we derived better bounds
by exploiting the symmetry of isotropic states \cite{WJD}. Apart
from their fundamental significance, our results are interesting
from the point of view of the characterization of quantum
information resources: if the noise affecting a state is larger
than our bounds, its outcome correlations for local measurements
can be reproduced by classical means alone.

\textit{Note added.} While completing this work, we learned that
our local model for isotropic states was independently derived in
\cite{WJD} in the context of state steerability. We note that all
our models imply the non-steerability of the corresponding quantum
states because Alice's response function is always quantum (see
\cite{WJD} for details).

\textit{Acknowledgements.} We acknowledge financial support from the
EU Qubit Applications Project (QAP) Contract number 015848, the
Spanish projects FIS2004-05639-C02-02, Consolider QOIT, the Spanish
MEC for ``Ramon y Cajal" and ``Juan de la Cierva" grants, the
Generalitat de Catalunya, the Funda\c{c}\~{a}o para a Ci\^{e}ncia e
a Tecnologia (Portugal) through the grant SFRH/BD/21915/2005, the
National Research Fund of Hungary  OTKA under Contract No. T049234,
and the Hungarian Academy of Sciences (Bolyai Programme). Research
at Perimeter Institute for Theoretical Physics is supported in part
by the Government of Canada through NSERC and by the Province of
Ontario through MRI.

\bibliographystyle{plain}

\end{document}